\documentclass{aa}
\usepackage{graphicx}
\usepackage{natbib}
\usepackage{txfonts}
\begin{document}
   \title{A heuristic derivation of the uncertainty of the frequency determination in time series data}

   \author{Thomas Kallinger,
   	  Piet Reegen
          \and
          Werner W. Weiss
          }

   \offprints{kallinger@astro.univie.ac.at}

   \institute{Institute for Astronomy (IfA), University of Vienna,
              T\"urkenschanzstrasse 17, A-1180 Vienna
             }

   \date{Accepted 22 December 2007}

   \abstract
   {Several approaches to estimate frequency, phase and amplitude errors in time series analyses were reported in the literature, but they are either time consuming to compute, grossly overestimating the error, or are based on empirically determined criteria.}
   {A simple, but realistic estimate of the frequency uncertainty in time series analyses.}
   {Synthetic data sets with mono- and multi--periodic harmonic signals and with randomly distributed amplitude, frequency and phase were generated and white noise added. We tried to recover the input parameters with classical Fourier techniques and investigated the error as a function of the relative level of noise,  signal and frequency difference.}
   {We present simple formulas for the upper limit of the amplitude, frequency and phase uncertainties in time--series analyses. We also demonstrate the possibility to detect frequencies which are separated by less than the classical frequency resolution and that the realistic frequency error is at least 4 times smaller than the classical frequency resolution.}
   {}
\authorrunning{Kallinger et al.}
   \maketitle
%
%________________________________________________________________

\section{Motivation}
In the frequency analysis of time series, a realistic estimate of the amplitude, phase and frequency uncertainties can be of special interest. Few examples are:
\begin{itemize}
\item The comparison of frequencies derived for simultaneously observed stars allows identifying instrumental signal, if the frequencies occur in different data sets but within the frequency uncertainty.
\item One needs to know the observed frequency errors in order to assess the quality of a fit of models to the observations.
\item For mode identifications based on amplitude ratios or phase differences from multi--color photometry one also needs a reliable estimate for the frequency error.
\end{itemize}

A combination of Fourier and least--squares fitting algorithms (like {\it SigSpec} by \citealt{Reegen}, {\it Period04} by \citealt{Lenz}, or {\it CAPER} by \citealt{Walker}) is a frequently used method for determining frequencies, amplitudes and phases of harmonic signals. For a time series consisting of a perfect sine wave and white noise, the frequency error is determined by the total time base of the data set and the signal--to--noise ratio ($SNR$) of the corresponding amplitude in the Fourier spectrum. \cite{Montgomery} defined the amplitude, phase and frequency errors as 
   \begin{equation}
   \sigma(a)_{\mathrm{Montgomery}} = \sqrt{\frac{2}{N}} \sigma (m),
   \label{EqAmp}
   \end{equation}
   \begin{equation}
   \sigma(\phi)_{\mathrm{Montgomery}} = \sqrt{\frac{2}{N}} \frac{\sigma (m)}{a},
   \label{EqPha_M}
   \end{equation}
   \begin{equation}
   \sigma(f)_{\mathrm{Montgomery}} = \sqrt{\frac{6}{N}} \frac{1}{\pi T} \frac{\sigma (m)}{a},
         \label{EqFerr_M}
   \end{equation}
based on an analytical solution for the one--sigma error of a least--squares sinusoidal fit with a rms of $\sigma (m)$. The total number of data points, the total time base of the observations, the signal amplitude, phase and frequency are $N$, $T$, $a$, $\phi$ and $f$, respectively. Hence the last term in Eq.\,\ref{EqPha_M} and \ref{EqFerr_M} represents SNR$^{-1}$ in the time domain. We want to mention that the time domain SNR in these relations is not equal to the commonly used SNR in the amplitude spectrum (peak amplitude divided by the average amplitude in a given frequency range) and it scales to the time domain SNR by a factor of $\approx \sqrt{\pi / N}$.    \cite{Reegen} shows that this scaling cannot be applied uniquely to the full frequency range and that systematic effects have to be taken into account if an exact description of frequency--domain errors is needed.

However, in reality an intrinsic signal is superposed not only by white noise (e.g. due to photon statistics) but also by correlated noise (e.g. atmospheric scintillation for ground--based data) or non--Gaussian distributed noise (e.g., introduced by the data reduction). Even the star itself can contribute correlated noise, for example due to granulation. All these noise sources increase the real frequency uncertainty which leads to the unsatisfying situation that in the literature several empirical parameters can be found which tune the frequency error to personal experience. 

People quite often use the Rayleigh frequency resolution ($T^{-1}$), defined by the total time base of the data set, which is in most cases a dramatic overestimation of the real uncertainty.  
To access the uncertainties of the fitting parameters for the time series analysis, it turned out to be an appropriate way to perform simulations with the actually analyzed data set, as it is done by Monte Carlo simulations in {\it Period04} or by bootstrap simulations in {\it CAPER} (see \citealt{Rowe} for details). This approach has the disadvantage that the simulations can be very time consuming especially if the data sets are big and/or include plenty of signal components.

%______________________________________________ Fig. Ferr
   \begin{figure}[th]
   \centering
      \includegraphics[width=0.5\textwidth]{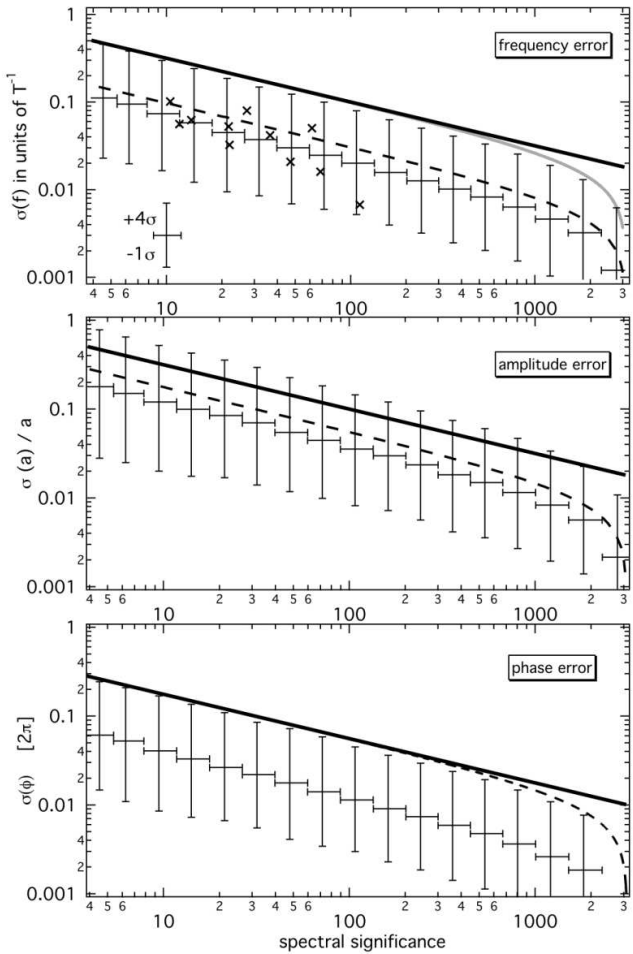}
      \caption{
      {\it top:} Frequency error $\sigma(f)$ normalized to the Rayleigh frequency resolution given by the data set length $T$ versus the spectral significance. Given are average values in bins (represented by the horizontal bars) as a result of a numerical simulation of 42\,597 synthetic data sets including a single sinusoidal signal with random frequency, amplitude, phase and white noise added.  Vertical bars indicate the +4$\sigma$ (and -1$\sigma$) distribution illustrating that the heuristically determined frequency error criterion (solid black line) represents a good approximation for the upper limit of the frequency uncertainty which is at least by a factor significance$^{1/2}$ smaller than the frequency resolution $T^{-1}$.
Cross symbols correspond to frequency errors derived from the comparison of real ground based data with high--precession space photometry of the same stars. For an explanation of the grey line see last but one paragraph of Sec. 2.1.
{\it middle:} Relative amplitude error versus the spectral significance. The solid line indicates the upper limit for the relative amplitude error given in this work.
{\it bottom:} Phase error (in units of $2\pi$) versus the spectral significance. The solid line shows the 
``Montgomery phase error" converted to spectral significance. {\it all panels:} The dashed lines represent the analytically determined one sigma error of a sinusoidal least--squares fit (Montgomery \& O'Donoghue 1999).
}
         \label{FigFErrS}
   \end{figure}

\section{Mono-periodic signal}

To quantify the effect of white noise on the frequency determination of a coherent mono--periodic signal, a numerical simulation was performed for 42\,597 synthetic data sets. 
Each data set consists of 10\,000 data points uniformly distributed over 10 days and includes two components: a single sinusoidal signal with random (uniformly distributed) frequency, amplitude and phase, and Gaussian distributed scatter with a random (uniformly distributed) amplitude (FWHM of the Gaussian random--number generator). All input parameters are independent of each other.

\subsection{Frequency error}

For the frequency analysis, the routine {\it SigSpec}\footnote{{\bf Sig}nificance {\bf Spec}trum,\\ http://www.astro.univie.ac.at/SigSpec/} (\citealt{Reegen}) was used. It is an automatic program to detect periodic signals in data sets and relies on an exact analytical solution for the probability that a given DFT (Discrete Fourier Transform; \citealt{Deeming}) amplitude is generated by white noise. Its main advantage to commonly used signal-to-noise ratio estimates is its appropriately incorporated frequency {\em and} phase angle in Fourier space, and time--domain sampling, hence using all available information instead of mean amplitude only. 
The {\it SigSpec} spectral significance is defined as the logarithm of the inverse False--Alarm Probability that a DFT peak of a given amplitude arise from pure noise in a non--equidistantly spaced data set.

On average, a SNR of 4 corresponds to a spectral significance value of 5.46. This means that an amplitude of four times the noise level would appear by chance at a given frequency in one out of $10^{5.46}$ cases, assuming white noise.

Fig.\,\ref{FigFErrS} shows the absolute deviation -- scaled to the data set length -- between the input frequency and the {\it SigSpec} frequency as a function of the spectral significance. Given are average values in bins of spectral significance (indicated by the horizontal bars). Not surprisingly, there is a clear dependency of the frequency error on the significance (or SNR). Vertical bars indicate the +4$\sigma$ (and -1$\sigma$) distribution of our simulation. Obviously, the real frequency error quite often ($\approx$ 30\,\%) exceeds the frequency error given by Eq.\,\ref{EqFerr_M} and which is indicated by a dashed line in Fig.\,\ref{FigFErrS}. 
However, we could heuristically define a frequency error criterion (solid black line in the top panel of Fig.\,\ref{FigFErrS}) as

   \begin{equation}
   \sigma(f)_{\mathrm{Ka}} = \frac{1}{T \cdot \sqrt{sig}} \approx \frac{\pi \cdot \mathrm{log}e}{4 \cdot T \cdot SNR},
   \label{EqFerr_K}
   \end{equation}
representing a good approximation for the upper limit of the frequency uncertainty and showing that the frequency uncertainty is less than the frequency resolution $T^{-1}$, at least by a factor of $\sqrt{sig}$. 
Only 4 out of 42\,597 simulations result in a frequency error exceeding the so defined upper frequency error limit. Being aware that a simulation need not reflect the reality, we added the frequency error of real observations into Fig.\,\ref{FigFErrS} (large dots) derived from the comparison of ground based data with long--term high--precision space photometry (MOST) of the same stars. 
We have to mention that plotting the frequency error as a function of the signal frequency (or phase) reveals no correlation between these quantities (in order to be independent from the spectral significance, synthetic data sets with a fixed SNR have been used).

\begin{figure}[t]
   \centering
   \includegraphics[width=0.5\textwidth]{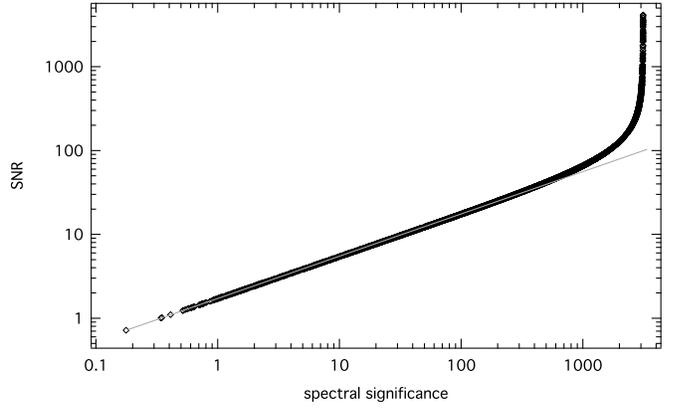}
      \caption{Amplitude spectrum signal-to-noise ratio (SNR) versus spectral significance for frequencies determined from 42\,597 synthetic data sets. The deviation from the linear relation (gray line in the log-log plot) at high significances is due to different noise estimate for SNR and spectral significance. Whereas the SNR is based on the rms scatter of the time series {\it after} prewithening the signal, the significance is based on the rms scatter of the time series {\it including} the signal.
             }
         \label{FigSNSig}
   \end{figure}

The deviation from a linear relation at high significances in the log--log scale of Fig.\,\ref{FigFErrS} is due to a distortion of the significance scale which is explained in Fig.\,\ref{FigSNSig}, where the SNR in the amplitude spectrum is plotted versus the spectral significance for frequencies determined from the synthetic data sets. 
For spectral significances below some hundred the significance is roughly equal to $(\pi \cdot \mathrm{log\,e}) / 4$ times the SNR$^2$ in the amplitude spectrum (\citealt{Reegen}). 
Only for extremely significant signals one has to take into account that the noise calculation for the SNR and the spectral significance is different. Whereas the SNR is based on the average amplitude in a Fourier spectrum {\it after} prewithening the signal (corresponds to the rms residual), the spectral significance is based on the rms scatter of the time series {\it including} the signal. With other words, a pure signal without noise has an infinite SNR but still a finite spectral significance (see \citealt{Reegen} for details).
The grey line in Fig.\,\ref{FigFErrS} takes this effect into account.

In order to explain the difference between the upper frequency error limit and the ``Montgomery frequency error", we interpret the latter to be the statistically expected value for the frequency uncertainty corresponding to the average values in the spectral significance bins of our simulation. Finally, we have to mention that the frequency error distribution of our simulation (for fixed spectral significance) is neither Gaussian nor symmetric which makes it very difficult to define an analytical average value and scatter for the frequency uncertainty.

\subsection{Amplitude error}
Whereas the absolute amplitude error only depends on the time series rms scatter (see Eq.\,\ref{EqAmp}), the relative amplitude error $\frac{\sigma(a)}{a}$ should be correlated with the signal's spectral significance (or SNR). The middle panel of Fig.\,\ref{FigFErrS} shows the relative amplitude error (deviation between the input amplitude and the {\it SigSpec} amplitude relative to the {\it SigSpec} amplitude) versus the spectral significance of our simulated white noise data sets. The dashed line indicates the relative amplitude error based on the absolute ``Montgomery amplitude error" representing the statistically expected value. According to our upper limit for the frequency uncertainty, we could again define an upper limit for the amplitude error of a sinusoidal least--squares fit as follows,

   \begin{equation}
   \frac{\sigma(a)_{\mathrm{Ka}}}{a} = \frac{1}{\sqrt{sig}} \approx \frac{2}{\sqrt{\pi \cdot \mathrm{log}e}} \cdot \frac{1}{SNR},
   \label{EqAerr_K}
   \end{equation}
indicated as solid line in the middle panel of Fig.\,\ref{FigFErrS}. However, the upper limit for the amplitude error is not as good defined as for the frequency error. But still $\approx$ 98\% of the determined amplitude errors are smaller than the given limit. 

\subsection{Phase error}

The bottom panel of Fig.\,\ref{FigFErrS} illustrates the absolute deviation between the input phase and the {\it SigSpec} phase versus the spectral significance of the 42\,597 synthetic data sets. Again, the dashed line indicates the phase error for a sinusoidal least--squares fit according to Eq.\,\ref{EqPha_M}. Contrary to the ``Montgomery frequency error" corresponding to the statistically expected value for the frequency uncertainty, the ``Montgomery phase error" is consistent with an upper limit for the real phase error. All, but 4 numerically determined phase errors are below the given limit.
Eq.\,\ref{EqPha_M} based on the time--domain SNR is converted to spectral significances (and frequency--domain SNR) 
as follows,

   \begin{equation}
   \sigma(\phi) = \sqrt{\frac{\mathrm{log}e}{2 \cdot sig}} \approx \sqrt{\frac{2}{\pi}} \cdot \frac{1}{SNR},
   \end{equation}
which is indicated by a solid and a dashed line in the bottom panel of Fig.\,\ref{FigFErrS}.

%______________________________________________ Fig. SN vs. Sig

\section{Multi-periodic signal}
Usually the smallest frequency separation of two independent signals in a data set which can be determined separately is called frequency resolution. 

For two signals with comparable amplitudes, a frequency separation corresponding to the Rayleigh frequency resolution (T$^{-1}$) results in a local minimum between the two peaks in the amplitude spectrum.
Closer frequencies produce an asymmetric peak whereas the peak maximum is roughly at the amplitude--weighted mean of the frequencies. 
After prewithening the signal (corresponding to the subtraction of a scaled spectral window at the given frequency, \citealt{Roberts}) some signal will be still left in the amplitude spectrum. With other words, it should be possible to determine frequency, amplitude and phase of signals separated in frequency by less than the frequency resolution. Hence, the uncertainties of these parameters should be less than given by the Raleigh criterion.

To quantify this uncertainty, a numerical simulation was performed for $\sim$50\,000 synthetic data sets now including two signals with random frequency, amplitude and phase for the first component. The second signal has a frequency randomly separated from the first one between 0 and 5 times the Rayleigh frequency resolution ($T^{-1}$), a random amplitude between 0.1 and 1 times the amplitude of the first one and a random phase. Gaussian distributed scatter with a random amplitude was added to the synthetic data.

\begin{figure}
   \centering
   \includegraphics[width=0.5\textwidth]{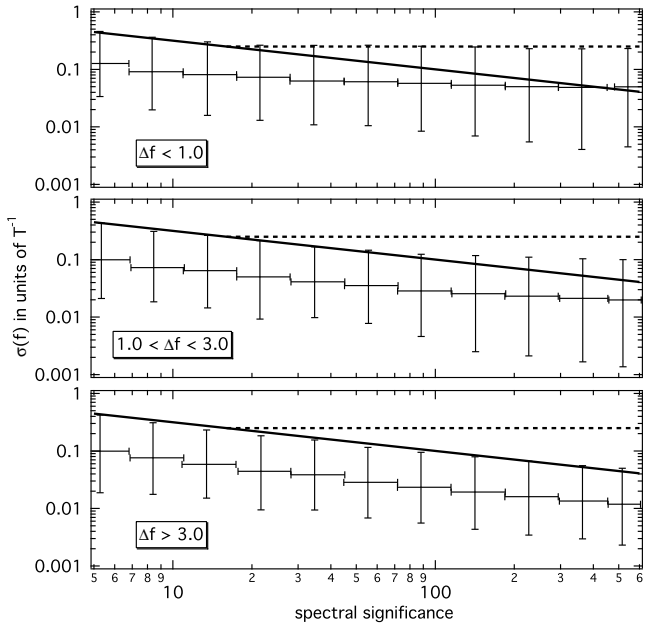}
      \caption{Same as top panel of Fig.\,\ref{FigFErrS} now including two sinusoidal signals illustrating average frequency errors $\sigma(f)$ (normalized to the Rayleigh frequency resolution) of the stronger signal (first detected in the prewhitening sequence) in bins of the spectral significance along with +4$\sigma$ (and -1$\sigma$) environments in the bins. The panels refer to different ranges of the frequency separation $\Delta f$ (in units of the $T^{-1}$) of the two input signals. The solid line indicates the upper frequency error limit for mono--periodic signals.
The dashed line corresponds to the heuristically determined upper frequency error limit for close frequencies and is equal to $(4 \cdot T)^{-1}$.
             }
         \label{FigFErrM}
   \end{figure}

Fig.\,\ref{FigFErrM} shows the average absolute frequency error in bins of the spectral significance of the stronger signal for different ranges of the frequency separation $\Delta f$ (in units of the Rayleigh frequency resolution) of the two input signals. 
The presence of a second signal separated by less than the Rayleigh frequency resolution limits the frequency uncertainty of the stronger signal to $(4 \cdot T)^{-1}$ (see dashed lines in Fig.\,\ref{FigFErrM}) if the spectral significance exceeds a value of 16 (this is where both criteria give the same frequency error). We have to note that this limit is again purely heuristically determined. 
For a second signal, separated by more than 3 times the Rayleigh frequency resolution, the frequency uncertainty of the stronger signal is limited by the frequency error criterion for a mono--periodic signal given by Eq.\,\ref{EqFerr_K} (see bottom panel in Fig.\,\ref{FigFErrM}). There seems to be a smooth transition for \makebox{1 $<$ $\Delta f$ $<$ 3} (middle panel).

Remarkably, only 13 out of $\sim$50\,000 ($\approx$ 0.026\,\%) numerically determined frequency errors do not satisfy the following criterion.

If a second signal is present within about three times the Rayleigh frequency resolution and spectral significance $>$ 16 the upper limit for the frequency error is  
	\begin{equation}
  	 \sigma(f)_{\mathrm{Ka}} = \frac{1}{4T}.
	\end{equation}
In all other cases the frequency error is smaller than
  	 \begin{equation}
  	 \sigma(f)_{\mathrm{Ka}} = \frac{1}{T \cdot \sqrt{sig}}
   	\end{equation}
corresponding to Equ. (4).

%______________________________________________ Fig. Multi Ferr

\section{Conclusions}
Based on extensive simulations, we have shown that there is an upper limit for the amplitude and frequency error in time series data analyses. Compared to the statistically {\em expected} value for the uncertainties given by \cite{Montgomery}, our {\em upper limits} cover the possible error due to white noise and leaves even room for additional error sources like atmospheric scintillation. 
A major advantage of calculating amplitude, frequency and phase errors in terms of spectral significance rather than signal--to--noise ratio is that the time--domain noise need not be Gaussian. As pointed out by \cite{Reegen}, the spectral significance does not depend on the probability distribution associated to the noise, and the only precondition is uncorrelatedness of consecutive data points. 
There has to be mentioned that amplitude, frequency and phase errors derived from spectral significances are only comparable to errors derived from SNR if the time series is well sampled (e.g. continuos space observations). Contrary to spectral significance based errors, SNR based error estimations (time--domain as well as frequency--domain) do not take into account the data sampling and can yield in a crude underestimation of the errors for ``bad" sampling like it is more or less always the case for single--site ground based observations.

We have shown that the phase error defined by \cite{Montgomery} is consistent with our simulations.

Furthermore we have shown that the determination of frequency pairs closer than the Rayleigh frequency resolution is possible and that the resulting frequency error is still 4 times smaller than the Rayleigh frequency resolution.
However, our simulation does not say anything about the reliability of close frequency pairs in general. It tells us about the frequency uncertainty of a peak if, after prewhitening this peak, a second significant peak is present. It tells us that peaks do not influence each other's frequency determination if they are separated in frequency by 3 times the Rayleigh frequency resolution. For closer peaks the frequency uncertainty is at least 4 times below the Rayleigh resolution even for peaks within the Rayleigh resolution.

\begin{acknowledgements}
This project was supported by the Austrian Fonds zur F\"orderung der wissenschaftlichen 
Forschung (FWF) within the project {\it The Core of the HR diagram} (P17580-N02), and the Bundesministerium 
f\"ur Verkehr, Innovation und Technologie (BM.VIT) via the Austrian Agentur f\"ur Luft- und Raumfahrt (FFG--ALR). 
\end{acknowledgements}


\begin{thebibliography}{}
  \bibitem[Deeming(1975)]{Deeming} Deeming T. J., 1975, Ap\&SS 36, 137
  \bibitem[Lenz \& Breger(2005)]{Lenz} Lenz P., Breger M. 2005, CoAst, 146, 53
  \bibitem[Montgomery \& O'Donoghue(1999)]{Montgomery} Montgomery M. H., O'Donoghue D. 1999, DSSN, 13, 28
  \bibitem[Reegen(2007)]{Reegen} Reegen P. 2007, A\&A, 467, 1353
  \bibitem[Roberts et al.(1987)]{Roberts} Roberts D. H., Lehar J., Dreher J. W. 1987, AJ, 93, 968
  \bibitem[Rowe et al.(2006)]{Rowe} Rowe J. F., Matthews J. M., Seager S., et al. 2006, ApJ, 646, 1241 %bootstrap
  \bibitem[Walker et al.(2005)]{Walker} Walker G., Kuschnig R., Matthews J. M., et al. 2005, ApJ, 635, L77 %CAPER
\end{thebibliography}
\end{document}